\documentclass[a4paper]{article}

\usepackage{INTERSPEECH2019}
\usepackage{hyperref}
\usepackage{subfigure}

\title{Towards joint sound scene and polyphonic sound event recognition}
\name{Helen L. Bear$^{1,2}$, In\^{e}s Nolasco{$^1$}, and Emmanouil Benetos$^{1,3}$\thanks{This work was funded under EPSRC grant EP/R01891X/1. EB is supported by a UK RAEng Research Fellowship (RF/128) and a Turing Fellowship. This work was supported by an NVIDIA GPU grant.}}
\address{$^1$School of Electronic Engineering and Computer Science, Queen Mary University of London, UK \\
$^2$Institute for Informatics, Technical University Munich, Germany\ \ \ \ \ 
$^3$The Alan Turing Institute, UK}
\email{\{h.bear,emmanouil.benetos\}@qmul.ac.uk, i.nolasco@se17.qmul.ac.uk}

\begin{document}

\maketitle
\begin{abstract}
Acoustic Scene Classification (ASC) and Sound Event Detection (SED) are two separate tasks in the field of computational sound scene analysis. In this work, we present a new dataset with both sound scene and sound event labels and use this to demonstrate a novel method for jointly classifying sound scenes and recognizing sound events.
We show that by taking a joint approach, learning is more efficient and whilst improvements are still needed for sound event detection, SED results are robust in a dataset where the sample distribution is skewed towards sound scenes. 
\end{abstract}

\section{Introduction}

\label{sec:intro}
Computational sound scene analysis refers to the field of study investigating computational models and methods for making sense of soundscapes in urban, domestic and nature environments \cite{CASSEbook}. Core problems in the field include identifying the acoustic environment of an audio stream, this is acoustic scene classification (ASC) or sound scene recognition \cite{barchiesi2015acoustic}, and on detecting the sound events or sound objects within a scene, namely sound event detection (SED) \cite{zhang2015robust}.
ASC and SED are commonly considered as two separate tasks in understanding sound scenes, as can be demonstrated by the evolution of the field through the IEEE AASP Challenges in Detection and Classification of Acoustic Scenes and Events (DCASE) \cite{stowell2015detection, mesaros2018detection, dcase2017}. 

\emph{Polyphonic} sound scenes are those containing multiple overlapping sound events, as opposed to \emph{monophonic} sound scenes that do not contain event overlaps \cite{stowell2015detection}. These can be background noise or foreground events where more than one sound source can be generating sounds at a single point in time. Polyphonic sound mixtures are challenging for recognising different sound sources and events, and this is the focus of this work.
It has been suggested that sound event information can help acoustic scene classification, meaning with event classifications a-priori, the accuracy of scene classification increases \cite{barchiesi2015acoustic}. Vice versa, with an accurate scene prediction the confidence of likely events in that scene increases \cite{heittola2013context}. In the latter case, SED can be described as \textit{scene-dependent} or \textit{scene-independent} \cite{cakir2017convolutional} but given the variability of events in a scene, this description does not work in reverse for ASC, i.e. that a scene is dependent or not on any single event.

Prior work includes training separate models for ASC and SED, where models require tuning for each task, for example \cite{phan2017improved}. Typically for ASC, researchers use Convolutional Neural Network (CNN) models (for examples we refer the reader to review the submissions online for the DCASE 2018 ASC task \cite{dcase2018}), since temporal dependencies are not considered important for ASC \cite{peltonen2002computational}. This differs to SED where researchers are most recently using Convolutional Recurrent Neural Networks (CRNNs) (for examples see Task 4 for \cite{dcase2018}) where local time information improves detection accuracy (e.g. \cite{cakir2017convolutional}). 

However, we are yet to predict both of these tasks simultaneously with a single model which is the main contribution of this paper. When recognising environmental sounds, humans use prior knowledge of likely events in the scene and their prior experiences of the scenes to classify environments, as demonstrated by one such listening test in \cite{chu2009environmental}. This, plus more evidence that context information such as a scene descriptor increases SED accuracy by machines \cite{heittola2013context}, motivates this work to build a single recognition model by learning both scene and event data concurrently. To the best of our knowledge this is the first attempt to create a system for joint ASC and SED. With respect to prior work in sound scene analysis, this is a novel proof-of-concept which has the potential to optimise future ASC and SED systems whereby robust ASC and SED inputs are coupled before training a single model to predict both scenes and events jointly. 
In the SED literature, detection and recognition refer to the same problem. This is because SED systems are not evaluated just on spotting events (with onsets and offsets), but on both spotting the events and assigning a label to them. For clarity, in the rest of this paper, we use the term \textit{classification} for matching class labels, \textit{detection} for spotting events and \textit{recognition} as both classification and detection.

\begin{table*}[t]
\centering
\caption{Proposed groupings of foreground sound event labels for each acoustic scene class.}
\begin{tabular}{l|l}
\toprule
Scene & Sound Events \\
\midrule
bus & clearthroat, cough, keys, laughter, phone, speech. \\
busystreet & bus-passby, doorclose, footsteps, key\_lock, knock, laughter, motorbike, speech, running, wind. \\
office &  chairs\_moving, doorslam, drawer, keys, knock, laughter, switch, phone.\\
openairmarket & bag\_rustle, bus-passby, cooking, footsteps, footsteps\_on\_grass, light\_rain, money, speech, wind.\\
park & bus\_passby, birdsong, footsteps\_on\_grass, gate, laughter, light\_rain, phone, pushbike, speech, wind. \\
quietstreet & birdsong, footsteps, key\_lock, light\_rain, pushbike, wind.\\
restaurant & chairs\_moving, cooking, doorclose, footsteps, laughter, speech. \\
supermarket & bag\_rustle, checkout beeps, footsteps, money, switch, trolley. \\
tube & announcement, bag\_rustle, footsteps, phone, slidingDoor\_close, speech, train.\\
tubestation & announcement, footsteps, running, slidingDoor\_close, speech, train.\\
\bottomrule
\end{tabular}
\label{tab:eventsInScenes}
\end{table*}

For ASC and SED, deep learning models based on CNNs/CRNNs are achieving good recognition accuracy \cite{mesaros2018detection, dcase2017, dcase2018}. A summary from the recent DCASE 2018 challenge submissions \cite{dcase2018} shows that for SED particularly, CRNN models are robust for sound event recognition. 
A CRNN is a neural network where the architecture includes both convolutional and recurrent layers \cite{goodfellow2016deep}. 
SED recognition scores are, as a rule of thumb, lower than ASC. However, it is an unfair comparison as ASC is usually treated as a single-label classification task, whereas SED is evaluated as a recognition task involving both detection and classification. Whilst SED can benefit from the recurrent layers in an CRNN, real-world applications of SED involve overlapping sound events which requires multi-label classification. Robust sound event features are the log of mel spectrogram energies \cite{lu2018multi} as this compact representation more closely approximates human perception, when compared with for example a magnitude STFT spectrogram.
Conversely for ASC, to predict one single label for an entire recording, the modeling of temporal dependencies is of lesser importance. Therefore, good features for ASC are temporally smoothed time-frequency representations (e.g. the approach in \cite{bisot2017feature}).

Whilst ASC and SED as separate tasks benefit from different models and inputs, real-world audio streams include both scene and sound event data. As humans listening to a real-world acoustic scene, one uses knowledge of a scene to help limit our choices of likely events, or if a distinct event is prevalent, known to only be likely in limited scenes, our expectation of predicting a specific scene increases. 
Benefits of a single model include: only one model to design and train; there is no integration or linear pipelining of separate models for each task; and by using data which contains only likely events in each scene, synthesized from real-world samples, it should generalize to real-life data by learning the variation of events from multiple different scenes, and thus we can predict both ASC and SED concurrently in the evaluation stage. 
The rest of this paper is as follows: the proposed joint SED and ASC method, including data curation, is described in Section~\ref{sec:method}. Results are presented in Section~\ref{sec:results} and conclusions are drawn in Section~\ref{sec:conclusions}.

\section{Method} \label{sec:method}

\subsection{Data preparation}
For the proposed model, a new dataset is needed as to the best of our knowledge all publicly available prior datasets are for a single task, meaning annotations for both sound scenes and events are not available. Recordings are designed such that the foreground sound event classes are only those likely to be in the background sound scene in the real world. For example, on a quiet street one might hear birdsong, but it is unlikely to hear the beeps from a supermarket checkout. Background scenes are taken from the DCASE 2013 ASC challenge private test dataset \cite{stowell2015detection}. For each of the ten scene classes, there are 20 recordings of 30 seconds long.
These recordings may contain unannotated foreground sound events as they are real-world recordings. So to mitigate the risk of erroneous false positives in evaluation, the background loudness in the final recordings is reduced. 

The first event samples are also taken from a DCASE challenge, this time from DCASE 2016 Task 2 (``Sound event detection in synthetic audio'', focusing on office sounds) \cite{mesaros2018detection}. Additional sounds are sourced from \texttt{FreeSound.org}. Only recordings of isolated sound events are used, not event sequences. Classes with greater intra-class variation (such as the \textit{footsteps} class in various shoe styles on different surface at different speeds has more samples than say, the \textit{wind} class which varies mostly by its strength) have a greater number of sources to redress variation imbalance.

\begin{table}[!ht]
\centering
\caption{Number of isolated event recordings per event class.  `*' classes are from DCASE 2016 Task 2; others are from \texttt{freesound.org}. \#R is the total number of recordings and \#T is the number of held-out test files.}
\resizebox{225pt}{!}{
\begin{tabular}{l|r|r||l|r|r}
\toprule
Sound Event & \#R & \#T & Sound Event & \#R & \#T\\
\midrule
announcement & 29 & 3 & bag\_rustle & 37 & 4 \\
birdsong & 39 & 4 & bus\_passby & 31 & 3\\
chairs\_moving & 35 & 3 & checkout\_beeps & 35 & 3\\
clear\_throat & 20 & 2 & cooking & 35 & 3\\
cough* & 20 & 2 & doorclose & 27 & 3\\
doorslam* & 20 & 2 & drawer* & 20 & 2 \\
footsteps & 30 & 3 & footsteps\_on\_grass & 36 & 4\\
gate & 30 & 3 & keys* & 20 & 2\\
key\_lock & 33 & 3 & lake & 38 & 4 \\
knock* & 20 & 2 & laughter* & 20 & 2\\
light\_rain & 28 & 3 & money & 34 & 3\\
motorbike & 29 & 3 & phone* & 20 & 2\\
pushbike & 30 & 3 & running & 20 & 2\\
sliding\_door\_close & 29 & 3 & speech* & 20 & 2\\
switch* & 20 & 2 & trolley & 28 & 3\\
train & 25 & 3 & wind & 36 & 4\\
\bottomrule
\end{tabular}
}
\label{tab:eventCounts}
\end{table}

All are recorded as \texttt{.wav} files, no re-encoded files from lossy compression formats are collected. Any sampling rates other than 44100Hz are upsampled or downsampled to match the scene sampling rate of 44100Hz. All files are max-normalised using python's \texttt{soundfile} library. Any leading silences are stripped. All source recordings are real-world ones. 

The total number of input event sources is tripled; for all foreground sound events a copy is produced with $+10dB$ relative to the source and a second duplicate with $-10dB$. This creates more variation in the synthetic scenes. 
The outcome of this sourcing is 824MB/5792.55sec (foreground) and 504MB/3000sec (background) data. 
For each of the 32 events there are at least 20 recordings from different sources. Certain DCASE 2016 event classes (alarm, keyboard, mouse, pen\_drop, and printer) were removed as they are specific to the office scene. The new events selected were based on likely events in each background scene as listed in Table~\ref{tab:eventsInScenes}, grouped by the paired background scenes.



As much as possible, event labels are in at least two different scenes to reduce the model learning `if event x occurs then it is always scene y'. This moves this work away from traditional closed set problems in ASC and SED towards real-world scenarios which are open set, thus an event can be occurring in multiple scenes \cite{bear2018}. The only exceptions to this are possible lake sounds in a park and trolley sounds in a supermarket. These are permitted due to the high likelihood of those sounds being associated  with the scenes but have very low likelihood of presence in the other scenes. 

To transform our collected sounds into scenes, \texttt{Scaper} \cite{salamon2017scaper} is used. \texttt{Scaper} is a python-based tool for synthesising sound scenes with accompanying annotation files. 
For each of our ten background scene classes, there are ten unique locations for each class which each having one sample recording, with a total of 100 background recordings. Each background is used in 10 new sound scenes for a new set of 1000 sound scenes.

During synthesis, foreground sound events are added. Parameters allow the event pitch to be altered by a random value between $-3$ to $+3$ semitones before it is added to the mixture and the event duration can be stretched with multipliers randomly selected between $0.8$ and $1.15$. The event-to-background signal-to-noise ratio (SNR) is randomly assigned in the range $-15$ to $15$ and all events for one scene are normally distributed throughout the 30 second scene. The number of events in a scene ranges from one to the total number of events in that background class plus one multiplied by three (e.g. scene class \textit{bus} has six possible event types so its event range is one to 21, likewise, scene class office has eight possible events, so ranges from one to 27 events in a 30 second recording). Thus scenes with more options are more likely to be busier scenes. Each scene is permitted duplicate event classes, and duplicate events from the same source for maximum variability. Event polyphony level in all scenes is three, i.e. the maximum number of concurrent sound events within a 30 second recording is three and event choice is random from the scenes list of event options. 
Next, all resulting sound scenes are augmented with pitch shifting using the MuDA Python library \cite{mcfee2015_augmentation}. In music signal analysis applications, this shift is often one or two semitones (e.g. \cite{tzanetakis2002musical}). Listening tests on the new sound scenes found that pitch could shift by six semitones up and down and the resulting scene remained realistic to human listeners. In order to create more variation in the dataset to support the recognition model to generalize well, each of the 1000 sound scenes are duplicated with two pitch shifts to treble the dataset, so the final sound scene and event dataset consists of 3000, 30 second \texttt{.wav} files with accompanying \texttt{JAMS} and annotation text files, a total size of $6.4$GB. All synthesised sound scenes are mono and available publicly\footnote{\url{http://doi.org/10.5281/zenodo.2565309}}.

\subsection{Feature extraction}
Features are extracted for every sound scene created as the log of mel-spectrogram energies using 128 mel bands, hop length of 512, and an STFT size of 2048.
Five folds of data are structured as per the background scene divisions from DCASE 2013 to ensure scene sources are not duplicated between train and test folds. After dividing the data into training and test folds, for each data sample we subtract the training fold mean and divide by the standard deviation for the fold. 
Next, a copy of each sound scene feature is smoothed over time as per \cite{bisot2017feature}, as this has been shown to be effective in ASC. Thus, the feature set contains one log-mel-specctrogram (useful for SED) and one temporally smoothed log-mel-spectrogram (useful for ASC).  
The final step stacks each 2D event feature with the 2D scene feature for each original recording into a 3D input such that the final training data are all two-channel inputs. 

In summary we have 1292 frames per 30second recording, 2100 training, 300 validation, and 600 test samples per fold. 


\subsection{Sound scene and sound event recognition}
Recognition is undertaken by first dividing the 3000 feature files into 600 for test, 2100 for training, and of the training files 300 are held out for validation using stratified five-fold cross validation.
Meaning, 20\% of data is for each test fold, and a quarter of the training data is held out for CRNN validation during training. There is no source/device/scene location/event overlap between train and test folds, consistent with the divisions in the original 2013 DCASE ASC task.   
The CRNN model architecture is three convolutional layers with max pooling, batch normalization, and dropout layers, followed by an LSTM layer, a fully connected layer, a further batch normalization and finally a Sigmoid activation layer in place of Softmax for multi-label classification with a binary-cross entropy loss function. 
Model parameters are detailed in Table~\ref{tab:params}. For the ASC baseline the class predictions are binarised with a global threshold of $0.9$ before majority voting (as per prior ASC work such as \cite{nguyen2018acoustic}). For the SED baseline we use the sed\_eval toolkit \cite{mesaros2016metrics} to measure the segment-based error rates (ER) and F1 score. The baseline models are ASC and SED models trained separately as single tasks against which we can compare the the proposed joint network. All are implemented in Tensorflow with Keras. For the joint recognition task we combine the measuring requirements with a minimum global threshold, tuned for each ASC/SED task. 

\begin{table}[!ht]
    \centering
    \caption{CRNN structure and parameters. Adam optimiser is used with $LR=0.001$, $beta_1=0.9$, $beta_2=0.999$, $epsilon=None$, $decay=0.0$, $amsgrad=False$.}
    \resizebox{\columnwidth}{!}{%
    \begin{tabular}{r|l}
    \hline
    Layer & params \\
    \hline \hline
    Convolutional & filters=64, kernel=(3,3)   \\
    MaxPooling2D & pool\_size=(3,3), strides=2, padding=`same' \\
    BatchNormalization & \\
    Dropout & prob\_drop\_conv=0.25 \\
    Convolutional & filters=128,  kernel=(3,3)   \\
    MaxPooling2D & pool\_size=(3,3), strides=2, padding=`same' \\
    Dropout & prob\_drop\_conv=0.25 \\
    Convolutional & filters=256, kernal=(2,2)   \\
    MaxPooling2D & pool\_size=(2,2), strides=2, padding=`same' \\
    BatchNormalization &  \\
    Dropout & prob\_drop\_conv=0.25 \\
    Reshape & shape=(256,-1) \\
    LSTM & filters=256, input\_shape=(1292, 128,channels) \\
    Dense & filters=256, activation=`relu'  \\
    Dropout & prob\_drop\_hidden=0.5  \\
    BatchNormalization & \\
    Dense & nb\_classes, activation=`sigmoid' \\
     \hline
    \end{tabular}}%
    \label{tab:params}
\end{table}

\begin{figure*}[!ht]%
\centering
\subfigure[ASC ground truth (first ten classes of joint task) ]{%
\label{fig:c}%
\includegraphics[width=\columnwidth,keepaspectratio]{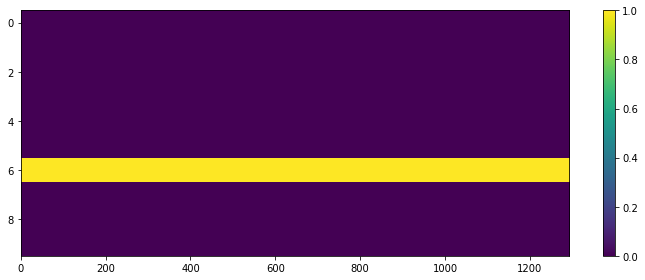}}%
\qquad
\subfigure[SED ground truth (final 32 classes of joint task).]{%
\label{fig:d}%
\includegraphics[width=\columnwidth,keepaspectratio]{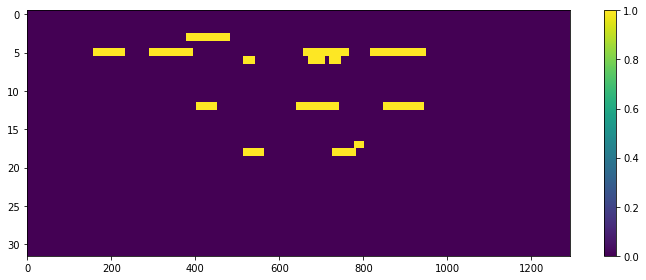}}%
\qquad
\subfigure[ASC predictions.]{%
\label{fig:e}%
\includegraphics[width=\columnwidth,keepaspectratio]{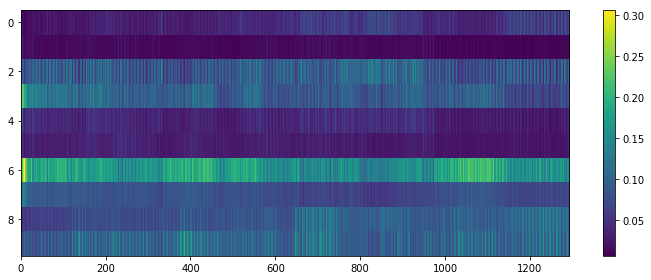}}%
\qquad
\subfigure[SED predictions.]{%
\label{fig:f}%
\includegraphics[width=\columnwidth,keepaspectratio]{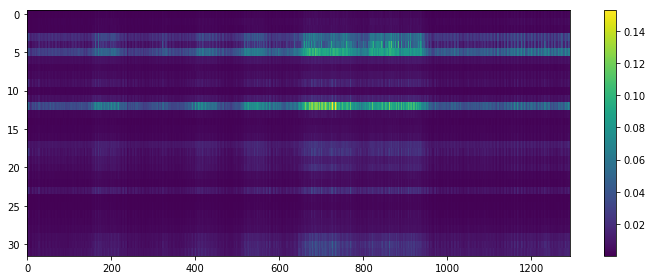}}%
\caption{An example test sample from source to prediction.}
\label{fig:eg}
\end{figure*}

Each feature vector input is the extracted feature matrices from each recording. All feature matrices are of equal length (one column for each time point) by 128 feature parameters. The corresponding label for each column is a $N$-hot vector of 43 binary values. 
The first ten elements represent scene classes and the remaining 32 are the sound event classes. 

\section{Results} \label{sec:results}
Conventional metrics for sound event detection (SED) are the segment-based Precision, Recall, the harmonic mean of these, the F1 score, and the Error Rate \cite{mesaros2016metrics}. 
Predictions are in the form of a binary vector; 
for each frame of the test samples, we produce a one hot vector. 
Measuring ASC is straightforward using classification accuracy (Acc) on scenes with majority voting on all the frames grouped by test recording sample as per many prior works \cite{8521242}. 

Care is taken in selecting metrics for SED due to the class imbalance between events as selections are random per scene creation in \texttt{Scaper}. We use two metrics from the SED\_eval toolkit \cite{mesaros2016metrics}, segment-based Error Rate (ER) (default segment size of one second is used) and the F1 score, this is common practice for comparison with other systems such as in the DCASE 2017 task 3. 
All code for feature extraction, recognition, and evaluation tasks is online\footnote{\url{https://github.com/drylbear/jointASCandSED}}.

\begin{table}[!ht]
\centering
\caption{Results of separate ASC \& SED models, compared with the proposed joint task model. MV = Majority Voting.}
\resizebox{225pt}{!}{
\begin{tabular}{l|r|r|r}
\hline
	&	 ASC 	& \multicolumn{2}{|c}{SED}  \\
        & MV Acc		& F1 	& ER  \\
    \hline \hline
    Separate models & $0.99\sigma0.01$   & $28.86\%\sigma2.67$  & $0.86\sigma0.01$\\
    Joint model    &  $0.98\sigma0.03$     & $13.73\%\sigma8.29$ & $1.00\sigma0.05$\\
    \hline 
\end{tabular}
}
\label{tab:results}
\end{table}

All results for ASC and SED are presented in Table~\ref{tab:results} and in Fig~\ref{fig:eg} there are four plots for an example test sample. The respective ground truths are shown in Figs~\ref{fig:eg}a and b for the ten ASC and 32 SED classes respectively and Figs~\ref{fig:eg}c and d show the predictions for this test sample with the joint-task model. The ASC prediction is distinctive but in this complex, polyphonic test sample, we see that the SED predictions are strong for timings (shown by lighter coloured frames), but where there are more than three classes in the ground truth, these are not so well detected. Also short event predictions are darker in Fig~\ref{fig:eg}d. 

First results in Table~\ref{tab:results} show robust ASC scores which we attribute to the fact that our dataset is built of real-world recordings with synthesised additions of specific events.
With the joint model there was no significant variation in mean ASC performance. But the joint models (for all folds) trained in approximately half the epochs of the separate ASC task, which supports the hypothesis that scene-specific event classes help ASC. 

There is a decrease in SED F1 and Error Rate when comparing the joint model with the separate SED model; we attribute this decrease to the skew in training samples per class once scenes and events are trained jointly. For example, our model is trained with per frame labels thus there are $30\times1292=38760$ training samples per scene, yet some event classes have as few as 140 training samples due to their short duration and sporadic appearance in each scene. We also attribute the greater standard deviation to this skew for the joint model SED F1, shown in Table~\ref{tab:results}. 

Reviewing class based measures, the events which perform poorly are either very short (e.g. door slam) or could be part of the background noise (e.g. motorbike). It is possible to artificially inflate the SED scores by increasing the segment-size (to improve detection accuracy by increasing recall, whilst risking extra insertion errors) or alter the prediction threshold for binarisation, but neither of these approaches gain more meaningful results thus we present the realistic SED results.
These results demonstrate the efficacy of the proposed joint model by comparison to separate ASC and SED tasks. To optimise the joint approach further one seeks a number of training epochs which does not overfit ASC but improves SED further than presented here. 








\section{Conclusions}
\label{sec:conclusions}
This paper presents a novel approach to jointly perform ASC and SED with a single model. In doing so, the work has produced a new publicly available dataset for sound scene and sound event recognition, with related scene and event classes, as well as onset and offset annotations for sound events. 

The new novel joint recognition model harnesses learning from the best of inputs used in ASC and SED but developing a model that will generalise well to two related but distinct tasks is a difficult undertaking and here we have shown promising results. These results show that a joint model can learn more efficiently.  
It does create future work for SED. For example: discovery of the right number of training epochs to improve SED without overfitting for ASC, altering the feature inputs to highlight distinct sound events and, revising the model architecture, therefore there is scope to build on and improve this work. Model performance has been quantified using metrics as benchmarks for the researchers interested in ASC and SED. 

\bibliographystyle{IEEEtran}

\bibliography{mybib}

\end{document}